\documentstyle[12pt,aaspp4]{article}    %%2.11.97  Ap J version
\def\K{{}^\circ {\rm K}}
\def\mixl{\lambda_{\rm mix}}

\slugcomment{}
\lefthead{Kennedy \& Bludman}
\righthead{Kennedy \& Bludman}

\begin{document}

\begin{flushright}
UF-IFT-HEP-95-4\\ UPR-718-T\\ NSF-ITP-95-71\\ 
Original: October 1996\\ Revised: February 1997
\end{flushright}
\title{VARIATIONAL PRINCIPLES\\ FOR\\ STELLAR STRUCTURE}
\author{Dallas C. Kennedy}
\affil{University of Florida}
\authoraddr{Gainesville FL 32611 USA;\\ kennedy@phys.ufl.edu}
\and
\author{Sidney A. Bludman}
\affil{University of Pennsylvania}
\authoraddr{Philadelphia PA 19104 USA;\\ bludman@bludman.hep.upenn.edu}
\begin{abstract}
The four equations of stellar structure are reformulated as two alternate
pairs of variational principles.  Different thermodynamic representations
lead to the same hydromechanical equations, but the thermal equations
require, not the entropy, but the temperature as the thermal field variable.  
Our treatment emphasizes the hydrostatic energy and the entropy
{\it production rate} of luminosity produced and transported.
The conceptual and calculational advantages of integral
over differential formulations of stellar structure are discussed
along with the difficulties in describing stellar chemical evolution by 
variational principles.
\end{abstract}
\centerline{Published in {\bf the Astrophysical Journal}: 20 July 1997}

\section{Differential Equations of Stellar Structure }

\subsection{Static Stellar Structure as a Non-Equilibrium Steady State}

In thermostatic equilibrium (Reichl 1980; Callen 1985), general statistical
properties radically simplify a macroscopic description.  Non-equilibrium 
systems, however, generally require a microscopic kinetic theory. In special 
non-equilibrium
systems, however, it is possible to leave  the microscopic physics implicit in
statistical averages, and to proceed with a nominally macroscopic dynamics.  
Stable stars evolve quasi-statically through
well-defined spatial structures and temporal stages, so that a macroscopic 
non-equilibrium thermodynamics can be formulated.   In this paper, we
apply non-equilibrium thermodynamics to stellar
structure, as distinct from stellar evolution.

Macroscopic non-equilibrium thermodynamics began with the work of Rayleigh
(1877) and Onsager (1931a, 1931b) (see also Casimir 1945); 
a complete theory was proposed by Prigogine and his followers
which applies to an important but restricted class of systems (Prigogine 1945;
Davies 1962; De Groot and Mazur 1962).  Later
developments have both extended (Donnelly {\it et al.} 1966; Glansdorff and 
Prigogine 1971) and rivaled this classic work (Tisza 1966;
Truesdell 1969; Gyarmati 1970; Lavenda 1978).  For stars,
we may take a generalized Prigogine approach, assuming some type of local
statistical equilibrium holds and intensive thermodynamics parameters are
defined at least locally in space and time.  This assumption validates a
macroscopic approach.  

Non-equilibrium thermodynamics is typically described by
conjugate pairs of variables: {\it differences} in intensive parameters, the
{\it thermodynamic forces,} and macroscopic, extensive {\it thermodynamic 
fluxes.}  Usually one relates forces to fluxes as cause to effect
and  assumes a linear or quasi-linear relation between the two.  In this paper,
we describe a quasi-static star of steady luminosity as an open system, 
receiving energy from nuclear sources which is ultimately radiated out of the 
stellar surface.

For the mechanical and thermal structure of this non-equilibrium steady state 
(NESS), we present two different pairs of variational principles (equations
(3.5) and either (3.11) or (4.2)).
While the differential formulation of stellar structure integrates 
local quantities from point to point, either integral formulation directly 
starts with global properties, including, but not limited to, total mass, 
luminosity, and radius.  Iterative application of a mesh approximation to 
the {\it global} variational integrals is analogous to the Henyey or 
relaxation {\it differential} method 
(Henyey, Vardya, and Bodenheimer 1965; Kippenhahn and Weigert 1990).
However, such a 
discretization approximation is not necessary, and continuous analytic 
approximations are also possible, in terms of variational parameters
of global significance. 

\subsection{Formal Similarity Between Mechanical and Thermal Equations}

The four first-order differential equations of quasi-static stellar 
structure occur in
two pairs: hydromechanical and thermal (Kippenhahn and Weigert 1990;
Hansen and Kawaler 1994).  In the Euler representation
and assuming spherical symmetry and conductive transport of luminosity,
\begin{equation}
\begin{array}{c}
%%-{dP/dr} = \rho{Gm\/r^2 + d^2 r/dt^2}\quad ,\quad {dm/dr} = 4\pi r^2\rho
%%\quad,\\
-{dP/dr} = {Gm\rho /r^2}\quad ,\quad {dm/dr} = 4\pi r^2\rho\quad, \nonumber \\
%%-{kdT/dr =l/{4\pi r^2}}    \quad ,\quad {dl/dr}   = 4\pi 
%%r^2\varepsilon\rho\quad,\eqnum{1.1}
-{kdT/dr =l/{4\pi r^2}}    \quad ,\quad {dl/dr}   = 4\pi r^2\rho(\varepsilon-
\varepsilon_\nu )\quad ,\end{array}
\eqnum{1.1}
\end{equation}\noindent
where the first horizontal pair is the mechanical (density-pressure) and the
second the thermal (luminosity-temperature) equations.   In radiative 
transport, the
thermal conductivity $k$ can be replaced by an equivalent radiative diffusion
expression $k\rightarrow 4acT^3/3\kappa\rho ,$ where $\kappa$ is
the opacity of matter, $ac/4$ the Stefan-Boltzmann constant, and the 
Boltzmann constant $k_B$ is set to unity throughout.
(Convective forms of luminosity transport are discussed 
below.)  The variables
are: $r$ = distance from center; $m(r)$ = cumulative mass from center to
$r,$ $P(r) = P_m(r) + P_\gamma (r)$ = total pressure (matter and radiation),
$\rho (r)$ = matter density;
$l(r)$ = photon luminosity at radius $r,$ $T(r)$ =
common matter-photon temperature, 
$\varepsilon (r)$ = luminosity production per unit mass, 
$\varepsilon_\nu (r)$=neutrino luminosity per unit mass,
$G$ = Newton's gravitational constant.  The luminosity is constant outside 
regions of luminosity production, where $\varepsilon$ = 0.

Except in Section~5, we neglect the slow nuclear chemical {\it evolution}, so that the
quasi-static stellar {\it structure} is in a NESS.  Dropping the time 
derivatives, if the thermal conductivity were constant or $\propto T^{-2},$
equations (1.1-2) would be self-adjoint and would admit 
a variational principle formulation.  Because the conductivity or radiation
transport coefficients generally do depend on dependent variables, the thermal
equations are not self-adjoint.  A self-consistent variational 
principle is nevertheless still possible (section~3).

Except for this technical difference, the mechanical and thermal equations of quasi-static stellar 
structure are now formally symmetric: momentum transport and luminosity transport equations, 
luminosity conservation laws, transport coefficients and density  are analogous.  
In this section, we exploit this
structural similarity between mechanical and thermal equations, while still stressing
the physical differences between the mechanical NESS
and the steady photon entropy generation.
       
In the static Euler representation, the seven dependent variables are: $m,$ $P,$ 
$\rho ,$ $T,$ $l,$ $\kappa$, and $\varepsilon$. There are
four differential and three constitutive equations; seven equations in all.  
With four first-order differential 
equations, there are four boundary conditions which are usually taken
as $P(R)$ = 0 (defining the
surface $r$ = $R$), $m(0)$ = 0 (no mass singularity at the center), $m(R)$ =
$M$ (total mass of the star), and $T(R)$ = $T_{\rm eff}$ (surface temperature
and thus total luminosity).

Because the included mass is conserved and accumulates monotonically with 
increasing radius, a point transformation to $m$ as the
independent variable is possible.
In this Lagrange representation,  the position $r$ of
each mass shell $dm$ is now dependent, $r$ = $r(m).$  The structure equations
now become:
\begin{equation}
\begin{array}{c}
-{dP/dm}   = {Gm/{4\pi r^4} }     \quad ,\quad {dr/dm} = 1/{4\pi 
r^2\rho}\quad, \nonumber \\
 -k{dT/dm}=       l/{16\pi^2  r^4} \quad ,\quad {dl/dm} = \varepsilon - 
\varepsilon_\nu\quad .\end{array}
\eqnum{1.2}
\end{equation}\noindent
The boundary conditions in the Lagrange representation assume $M$ as given.  
Thus, a useful set of boundary conditions is $r(0)$ = 0, $P(M)$ = 0, $T(0)$ = 
$T_c,$ $P(0)$ = $P_c.$

\subsection{Differences Between Mechanical and Thermal Steady States}

The first 
and third equations in the Lagrange representation show one difference between 
matter and radiation and between mechanical and thermal structure.   The star 
is defined by the presence of opaque matter,
but, unlike $m$, luminosity $l$ is still a dependent variable.  Because it 
does not
increase outside the core, $l$ would be useless as an independent variable
in the outside regions which transport, but do not produce, luminosity.

A physically more important difference between matter and radiation arises 
because the matter 
is static so that matter thermodynamics enters only implicitly, through the 
equation of state $P_m$ = $P_m(\rho ,T),$ where the matter internal energy 
acts as a potential for the pressure.
The thermal structure is not static, but shows
a stationary flow of released nuclear energy from the 
core to the surface.
Luminosity transport is explicitly statistical and depends
explicitly on thermodynamic averages: the local temperature $T$ and
the opacity $\kappa (\rho ,T).$  This non-equilibrium luminosity 
transport derives from 
non-equilibrium statistical mechanics (Appendix~A).  In radiative 
transport, this steady flow expresses the balance between  the outward force 
of radiation and 
the resisting force of opacity (radiation friction):
\begin{equation}
-{dP_\gamma\over dr} = {\kappa\rho\over c}{l\over{4\pi r^2}}\quad .
\eqnum{1.3}
\end{equation}\noindent
Convection is more complex than radiative transport and usually described in 
mixing length theory by the mixing length
$\mixl$ as the transport parameter.  The heat flux becomes:
\begin{equation}
l/4\pi r^2 = (\rho c_PT)\mixl^2\sqrt{{\rm g}\gamma_P}~[{\nabla - 
\nabla_{\rm ad}
\over{\lambda_P}}]^{3/2}\quad ,\eqnum{1.4}
\end{equation}\noindent
for efficient convection, meaning that a convective cell loses little heat 
before it breaks up (Glansdorff and Prigogine 1971; Donnelly {\it et al.}
1966; Hansen and Kawaler 1994).  
(See Appendix~B for definitions.)   Convection is efficient in two cases:
The first is ``slow but
hot'', where $\nabla$ is only slightly larger than $\nabla_{\rm ad}$
(quasi-adiabatic regime), and the heat transport is efficient because the
heat capacity $c_P$ is large, but the cell velocity is slow.  This regime
occurs in the convective cores of stars with $M >1.08M_\odot .$  The
second case is ``warm but fast'', where $\nabla$ is well above
$\nabla_{\rm ad},$ $c_P$ is not large, but the cell 
velocity is high.  The outer convective layer of the Sun is in such a regime.

Convection only comes into play as a luminosity transport mechanism if the
Schwarzschild instability criterion holds.  Convection, radiative transport, 
and conduction
can operate simultaneously and in the same region of the star; in which  case,
the total luminosity is the sum of all three transport fluxes.  In the
case of efficient convective heat transport, virtually the entire luminosity 
flux is convective.

The non-equilibrium steady-state obtains in quasi-static stages of stellar 
evolution,
beginning with hydrogen and helium burning.  Steady states, after the
Main Sequence, burn
carbon, neon, oxygen, or silicon, and are dominated by neutrino production 
and transport.
The neutrinos escape essentially without matter interaction and without 
thermalizing.  
Neutrinos are produced by hot matter, but their
energy loss is not thermostatic: their luminosity contributes to the material 
internal
energy only by cooling the matter; they contribute nothing directly to stellar 
structure.  The neutrino
number and energy lost can only be inferred from the photon luminosity and
the assumed matter properties and nuclear reactions (Bahcall 1989; Kippenhahn 
and Weigert 1990; Hansen and Kawaler 1994; Arnett 1996).

\section{Stellar Structure as a Non-Equilibrium Steady State}

No set of differential equations has a unique variational formulation (Douglas
1941).
In this section and the next, we Legendre transform the global thermodynamic 
potentials,
leaving the Euler-Lagrange equations invariant. The energy representation we 
start with
enjoys a transparent physical interpretation in terms of
hydrostatic energy and entropy production.  In Section~4, we give
entirely
different variational principles in terms of local field variables, which 
nevertheless lead to the same mechanical and thermal equations (1.2).

\subsection{Thermodynamic Equilibria}

Thermodynamics distinguishes various types
of equilibria or steady states (Gyarmati 1970;
Glansdorff and Prigogine 1971; Reichl 1980; Callen 1985).
{\it Global} mechanical or thermal
equilibrium implies uniform pressure and temperature everywhere,
or at least throughout a large, finite system.  Stars are not in global
equilibrium; their pressures and temperatures vary in space, but
are still meaningful locally in small regions.  If the length scales over
which kinetic mechanisms maintain local mechanical or thermal 
equilibrium (LME or LTE) are
much smaller than the scale heights of pressure or 
temperature:
\begin{equation}
\lambda_P \equiv P/|{\bf\nabla}P|\quad ,\quad \lambda_T \equiv T/|{\bf\nabla}T|
\quad ,\eqnum{2.1}
\end{equation}\noindent
then a {\it local} pressure $P(r)$ or temperature $T(r)$ is well defined as a
function of $r.$  An
equilibrium can be {\it partial} or {\it complete} among various constituents 
of the star.  The LTE in a stellar interior is complete in that
all matter species and radiation have a common temperature $T_m$ =
$T_\gamma$ = $T.$  The pressure equilibrium is partial, in that the total
pressure is the sum of the partial pressures of each component,
$P$ = $P_m$ + $P_\gamma .$
The pressure decreases outwards because of the star's self-gravitation;
the temperature decreases outwards wherever the total luminosity 
exceeds the neutrino luminosity.

The local material internal energy density is determined by 
the matter pressure $P_m$. In LTE, $P_m$ is an equilibrium state
function of temperature and matter density, $P_m$ = $P_m(\rho ,T),$ the matter
equation of state.  The photon equation of state is a function of $T$ alone.
In the luminosity equations, the transport coefficient (the opacity or 
conductivity) and the specific luminosity production, in LTE, are 
functions of local state variables $T,~\rho,$ and~$X_i$, the local
chemical composition (nuclear mass abundances $X_i$).  The three 
constitutive functions 
depend on position only  implicitly, through the local
equilibrium variables $T(r),~\rho (r), X_i(r)$.  Since we are generally not 
considering
chemical evolution, we generally suppress the dependence on $X_i$.

Both LME and LTE break down in the stellar atmosphere,  where the matter 
becomes transparent and the radiation escapes almost unhindered.  Here matter 
and radiation are not in thermal equilibrium.
Pressure and gravity become unbalanced,
as the star emits a stellar wind into space (Chandrasekhar 1950; Stix 1989)

To distinguish these local thermodynamic equilibria from global
pressure and temperature equilibria, the global
state is referred to as a {\it non-equilibrium steady state} (NESS).  
Along the Main Sequence and any later, post-Main-Sequence steady states,
NESS obtains, with steady radial luminosity flow. 
Various time scales are associated with these local
equilibria and  NESS's, which  must be achieved and maintained by different  
mechanisms.  In the Sun,  LME and LTE are
achieved in about 10$^{-12}$ sec by local kinetic mechanisms.
The global mechanical and thermal
NESS's are established in about 30 min and 10$^{7}$ yrs, respectively.
These last two time scales characterize
helioseismological disturbances and non-static, macroscopic 
heat flows (Kelvin-Helmholtz
time).  Except in dynamic stages,
these time scales are much shorter than the local and global chemical 
evolution times 
10$^{7-8}$ and 10$^{10}$ yrs
associated with nuclear transmutations (Bahcall 1989; Kippenhahn and Weigert
1990; Hansen and Kawaler 1994; Arnett 1996).

\subsection{Different Thermodynamic Representations}

Like any thermodynamic system, a star can be described by different global
thermodynamic potentials.  The original representation is the {\it energy 
representation}
$E$ = $E(V,S,N_i),$ as a function of volume $V,$ entropy $S,$ and species
numbers $N_i.$  Because the thermodynamic state varies spatially, the
extensive state variables
must be locally recast, as either spatial (per unit volume) or
specific (per unit mass) densities (Chandrasekhar 1939; Kippenhahn and Weigert
1990).  (We follow astrophysical custom by
using the latter, unless otherwise noted.)  Extensive quantities are
then mass integrals over the specific densities.
The specific energy $e$ is the sum of the specific
internal energy $u$ and the specific gravitational energy
$-G m/r.$  The specific volume $v_\rho$ is $1/\rho .$  In all but the hottest 
stars,
the total entropy $S= S_m + S_\gamma$ is dominated by matter, with specific 
entropy 
$s_m.$  Nevertheless, the small radiation entropy {\it flux}
is responsible for the stellar luminosity and cannot be ignored.

In the stellar interior, where a NESS obtains, the specific entropy 
$s= s_m + s_\gamma $ is stationary,
so that $E=E(V,S)$ and $e(\rho , s_m + s_\gamma )$ are,
respectively, global and specific thermodynamic potentials.
The radiation entropy density alone is determined by the local ratio
$\kappa l/m.$
If the thermal structure of the star is prescribed 
(e.g., isentropic or isothermal), a barotropic relation $P=P(\rho)$ holds so that 
the hydromechanical equations are closed.  Nuclear burning in the stellar core 
leads to secular chemical changes (which we ignore in stellar structure), and to the rapid 
thermalization of fusion products, which steadily produces radiation entropy flux
out of the star.  In the steady state, the star's entropy is constant and cannot be used as 
a global thermodynamic state function.
In order to treat matter and radiation as symmetrically as possible, 
we  use the temperature $T,$ which is common for the matter and radiation
in the stellar interior.  The appropriate generalization for the stellar
atmosphere is straightforward.  (See Appendix~A, subsection~3.)

The thermodynamic potential for the
new variables $(V,T)$ is the Helmholtz free energy
\begin{equation}
F(V,T) = E(V,S) - TS\quad \eqnum{2.2}
\end{equation}\noindent
or specific Helmholtz free energy $f(\rho ,T).$ 
(We might also go over to completely intensive state variables
$(P,T)$ and the Gibbs free energy $G(P,T)$ = $F + PV$
as the thermodynamic potential.  The choice of thermodynamic potentials
is a matter of convenience; we prefer the Helmholtz free energy.) 

\section{Global Thermostatic Potentials and Entropy Production}

In this section, we denote functional variations by $\delta ,$
local thermodynamic state and  spatial changes
by ordinary differentials,  so that the spatial differential  
$dF$ = ${\bf\nabla}F\cdot d{\bf r}.$

\subsection{Mechanical Steady State}

In considering only the mechanical NESS, we can use any thermodynamic 
representation.  We can use the total
hydrostatic energy (Lamb 1945; Chiu 1968; Rosenbluth {\it et al.} 1973; 
Hansen and Kawaler 1994):
\begin{equation}
E(V,S) = \int_0^M dm\ e(\rho (r),s(r))\quad ,\quad  e = u - G m/r
\quad ,\eqnum{3.1}
\end{equation}\noindent
or the Helmholtz free energy
\begin{equation}   
F(V,T) = \int^M_0 dm\ f(\rho (r),T(r))\quad ,\quad  f = e - Ts\quad ,
\eqnum{3.2}
\end{equation}\noindent
as the thermodynamic potentials.  The specific internal energy and free energy
have variations
\begin{equation}
\delta e  = -P\delta (1/\rho ) + G m \delta r/r^2 + T \delta s\quad ,
\eqnum{3.3} 
\end{equation}\noindent
\begin{equation}
\delta f = -P\delta (1/\rho ) + G m \delta r/r^2 - s \delta T\quad ,
\eqnum{3.4}
\end{equation}\noindent
where $\delta (1/\rho )$ = $d(4\pi r^2\delta r)/dm.$

We could also use the global enthalpy 
\begin{displaymath}
H(P,S) = E(V,S) + \int dV\ P\quad ,\quad dH = VdP + TdS\quad ,
\end{displaymath}\noindent 
or the Gibbs free energy
\begin{displaymath}
G(P,T) = E(V,S) + \int dV\ P - \int dS\ T\quad ,\quad dG = VdP + SdT\quad ,
\end{displaymath}\noindent
as global potentials.
The hydrostatic equation~(1.2) is obtained by freezing the thermal structure,
taking adiabatic variations in $E$ or $H$ or isothermal variations in $F$ or 
$G$.  For either mechanical variation, $\delta P/\rho=\delta h$ or $\delta g$.
In all cases, the functional variation does not assume that the frozen thermal 
structure is in the NESS.  But the short mechanical time scale compared to the
thermal time scale assures that  LTE, although not necessary, is a sufficient 
condition for LME.  ``Isothermal'' here refers not to {\it spatially 
constant} temperature, but {\it unvaried} temperature {\it profile;} similarly,
``adiabatic'' refers to unvaried specific entropy profile.

The simplest example of the resulting Euler-Lagrange equations arises from
applying the variation~(3.3) to~(3.1), setting $\delta s$ = 0 (Chiu 1968).  
Using $\delta (1/\rho )$ = $d(4\pi r^2\delta r)/dm$ and integrating the
first term in~(3.3) by parts, one obtains
\begin{equation}
\delta E = \int_0^M dm\ \Bigl[ 4\pi r^2(dP/dm) + Gm/r^2\Bigr]\delta r = 0
\eqnum{3.5}
\quad ,
\end{equation}\noindent
ignoring the boundary condition terms.  Setting the integrand here to zero
for arbitrary $\delta r$ reproduces the hydrostatic NESS, the first equation
of~(1.2).  Alternatively, applying the variation~(3.4) to~(3.2) isolates the 
mechanical structure in the isothermal limit $\delta T$ = 0.  

\subsection{Entropy Production in the Thermal Steady State}

In our NESS, the entropy density of matter and of radiation are each stationary
and very disparate in magnitude.  Because matter and radiation share a common 
temperature, we use $T$ as the intensive thermodynamic variable.  Instead of 
the entropy, our variational principle for thermal NESS minimizes the {\it 
entropy production rate} $\Sigma \ge$ 0 (Prigogine 1945; Davies 1962; De Groot
and Mazur 1962; Gyarmati 1970; Donnelly {\it et al.} 1966; Glansdorff and 
Prigogine 1971).  A simple minimum entropy production variational principle 
applies to systems close to global equilibrium in the linear regime and to a 
special class of systems, those in the quasi-linear regime.  In the linear 
regime, transport coefficients are constant.    In the quasi-linear regime, 
LTE still obtains, but the constitutive functions depend only on the local 
state variables and not on their derivatives.  Because the stellar NESS is in 
the nonlinear regime, the thermal variational principle depends on two 
temperatures, $T,~T_{\ast}$, the first of which is dynamical, the second 
referring to the background matter.
  
The entropy production rate is bilinear, a product of thermodynamic forces and
fluxes.  For temperature variations, the thermodynamic force arises, not from 
$T$, but from differences of $1/T$, or from the gradient of $1/T$.  In the 
linear or quasi-linear regime, the luminosity flux is linear in  $\nabla (1/T).
$  This thermal gradient drives the flow of luminosity from point to point, 
for any of the three standard heat transport mechanisms.  

The temperature also occurs in other parts of the entropy production rate, but
only as a {\it local} state variable having nothing to do with heat transport.
We must distinguish this temperature, $T_*,$ from the temperature $T$ which
is subject to functional variations (Donnelly {\it et al.} 1966; Glansdorff
and Prigogine 1971).  With this 
distinction, we can write the entropy production of conductive transport:
\begin{equation}
\Sigma_{\rm cond} = \int dV\ {1\over 2}(kT^2)_*[{\bf\nabla}(1/T)]^2\quad ,
\eqnum{3.6}
\end{equation}\noindent
and of radiative diffusion as:
\begin{equation}
\Sigma_{\gamma\ {\rm diff}} = \int dV\ {1\over 2}({4acT^5\over
{3\kappa\rho}})_*[{\bf\nabla}(1/T)]^2\quad ,\eqnum{3.7}
\end{equation}\noindent
with details in Appendix~A (Chandrasekhar 1950; Essex 1984a, 1984b).
The entropy production of convective transport is presented in Appendix~B
(B.2).  Any transport entropy production is a measure of the efficiency of 
these different mechanisms for luminosity transport, the 
price paid for transporting the luminosity produced in the core of the star 
to the surface. 

The entropy production due to nuclear burning in the core is:
\begin{equation}
\Sigma_{\rm fusion} = [\varepsilon\rho]_*/T\quad .\eqnum{3.8}
\end{equation}
The bulk radiation entropy production rate is the 
sum of the luminosity transport and production terms~(3.6-8):
\begin{equation}
\begin{array}{ccl}
\Sigma_\gamma & = & \Sigma_{\gamma\ {\rm diff}} + \Sigma_{\rm fusion} 
\nonumber \\
              & = & \int dV\ (1/2)[4acT^5/(3\kappa\rho )]_*
  [{\bf\nabla}(1/T)]^2 + (\varepsilon\rho)_*/T\quad ,\end{array}
\eqnum{3.9}
\end{equation}\noindent
in case of radiative transport.

Where the radiation outstreams from the photosphere, at temperature 
$T_{\rm eff},$
\begin{equation}
\Sigma_{\rm bound} = {4\over 3}(ac/4)T^3_{\rm eff}(4\pi R^2)\quad 
\eqnum{3.10}
\end{equation}\noindent
is obtained by integrating the radiation entropy flux over  the surface of the
star.  Because the mechanical NESS is reached so much faster than the thermal 
NESS, the mechanical structure  must be the hydrostatic
NESS corresponding to a given thermal distribution, stellar radius $R$ 
and surface temperature $T_{\rm eff}$ (Sieniutycz and Berry 1989, 1991,
1992, 1993; Sieniutycz and Salamon 1990; Sieniutycz
1994).  This boundary entropy production is
held fixed by the boundary conditions and plays no dynamical role in the 
thermal variational principle.

\subsection{Minimum Entropy Production}

Our thermal variational principle requires minimizing the entropy production 
by varying 
$T\rightarrow T + \delta T,$ but not $T_*$, holding $M$ and $T_\star$ fixed 
(Donnelly {\it et al.} 1966).  
Only after the Euler-Lagrange equation is obtained, is $T_*$ set
equal to $T.$  For each of the transport entropy productions (3.6, 3.7, B.2),
the standard heat transport equations are obtained in this way.  For example,
the radiative diffusion case is derived from~(3.9):
\begin{equation}
\delta\Sigma_\gamma = \int dV\ \Bigl[ [{4acT^5\over{3\kappa\rho}}]_*
[{\bf\nabla}(1/T){\bf\cdot}{\bf\nabla}\delta (1/T)] - 
[\varepsilon\rho ]_*(\delta T/T^2)\Bigr] = 0\quad .\eqnum{3.11}
\end{equation}\noindent
Integrating the first term by parts and dropping the boundary terms, we set
$T_*$ to $T$ and the entire integrand to zero.  For arbitrary variations 
$\delta T,$ the result is radiative diffusion:
\begin{equation}
{d\over{dr}}\Bigl({{16\pi r^2acT^3}\over{3\kappa\rho}}\cdot{dT\over{dr}}
\Bigr) + 4\pi r^2\varepsilon\rho = 0\quad .\eqnum{3.12}
\end{equation}\noindent
For simplicity, only the radial dependence has been kept, and the neutrino 
luminosity $\varepsilon_\nu$ has been ignored.  This same result is obtained
by combining the thermal-luminosity pair of equations in~(1.1).

In practice,
an ansatz is made for $T_*(r)$ and for $T(r),$ with only the latter containing
variational parameters to be determined by minimizing the variational integral.
The procedure is iterative: after each step $T_*(r)$ is set equal to the 
$T(r)$ obtained at the previous step.  One may use a global mesh, 
analogous to that in the {\it differential}
Henyey method, or an analytic form with a finite or infinite number of 
adjustable and physically suggestive parameters.
The procedure is identical to the background field (BF)
or self-consistent field (SCF) method used in quantum many-body and field
systems and can lead to an analytic approximation for the stellar 
structure in terms of global parameters.  These may be related to global
properties and boundary conditions of the star.

Because no entropy production is associated with the thermalization or 
diffusion of neutrinos,  the only neutrino entropy production is by averaging 
over neutrino energies produced by the thermalized reactant matter. 
The exact
expression for neutrino entropy production varies by reaction, but for high
temperatures, its dimensional order of magnitude is very approximately:
\begin{displaymath}
\Sigma_\nu \sim {\dot N}_\nu\quad ,
\end{displaymath}\noindent
where ${\dot N}_\nu$ is the total neutrino production rate.  (A better 
estimate is given in Appendix~C, C.10)  The neutrino contribution 
to entropy production is usually much
smaller than the photonic.  But in advanced stages of stellar evolution,
the temperature is high enough to bring weak
interactions significantly into play, letting neutrino entropy production 
rival that of heat and radiative diffusion mechanisms.

In supernova explosions, neutrinos do play the
same role as photons in ordinary stars: they
interact with the ambient stellar matter, are thermalized, and are then
emitted from the surface of the  neutrinosphere at a definite 
blackbody temperature (Bahcall 1989; Arnett 1996). The temperature in the 
innermost region reaches 
a very high 30-50 MeV, so that neutrino luminosity transport is efficient and 
neutrino entropy production remains fairly small, below that of other
energy transport mechanisms.

\section{Variational Principles with Local Specific Potentials}

In the preceding section, we presented variational principles for the 
hydrostatic NESS (3.1-5) and for steady-state luminosity transport (3.6-8).
For mechanical NESS alone, any global thermodynamic potential can be 
extremized, but 
for the thermal steady state, only the global potentials, $G(P,T)$ or $F(V,T),$
and a close relative, $\Sigma (T),$  could be used.
Independently of thermodynamics, we know that the same differential
equations admit of many different variational integral forms.
Because  the specific potentials, potentials per unit mass,
vary in space and also represent the local thermodynamic state just as well as the 
variables $(V,P,T,S),$ we may take the specific potentials as
field variables and find other variational integrals to serve as
global actions.  

In the mechanical equations, we can switch from the density $\rho$ and pressure $P$ as
the intensive state variables to either the specific enthalpy
\begin{equation}
h(P,s) = e(\rho ,s) + P\rho\quad ,\quad dh = \rho~dP + Tds\eqnum{4.1a}
\end{equation}\noindent
or the specific Gibbs free energy
\begin{equation}
g(P,T) = e(\rho ,s) + P\rho - Ts\quad ,\quad dg = \rho~dP + sdT\quad ,
\eqnum{4.1b}
\end{equation}\noindent
as the new local field variables.
The mechanical NESS is obtained by either varying
\begin{equation}
{\cal I}_{\rm ad} = \int dV\ [-({1\over{8\pi G}})({\bf\nabla}h)^2 + P(h)]\quad
\eqnum{4.2a}
\end{equation}\noindent
adiabatically or 
\begin{equation}
{\cal I}^\prime_{\rm iso} = \int dV\ [-({1\over{8\pi G}})(\nabla g)^2 + P(g)]
\quad ,\eqnum{4.2b}
\end{equation}\noindent
isothermally, so that $\delta P/\rho$ = either $\delta h$ or $\delta g$.

For the thermal part, we define a {\it heat} or {\it radiation potential
density} $\theta ,$ where $d\theta$ = ${\bf J}_Q\cdot d{\bf r}$ and $d{\bf
r}$ is the distance increment through which the luminosity is transported.
In conductive or radiative transport,  $d\theta$ = $-k dT$ or 
$-(c/\kappa\rho )dP_\gamma .$  In convective transport,
\begin{equation}
d\theta = (\rho c_PT)(\mixl /\lambda_P)[\nabla - \nabla_{\rm ad}](w\ dr)
\quad ,\eqnum{4.3}
\end{equation}\noindent
where $w$ = upward speed of the convective cell, assuming $\nabla\ge
\nabla_{\rm ad}$ (see Appendix~C).  The luminosity source is included by a
{\it source potential} $\Pi (\theta ),$ where $d\Pi$ = 
$\rho \varepsilon d\theta .$  The thermal variational integral is then:
\begin{equation}
{\cal J} = \int dV\ [-({1\over{2c}})(\nabla\theta )^2 - \Pi (\theta )]\quad ,
\eqnum{4.4}
\end{equation}\noindent
where $r$ is the independent variable and $\theta$ the field.  Variation
of~(4.4), keeping the mechanical structure fixed, yields the thermal 
transport equations for all three kinds of transport.  These actions 
${\cal I}, {\cal J}$ do not have the simple thermodynamic
interpretation that the minimal entropy production principle enjoys.

\section{Variational Thermohydrodynamics and Chemical Evolution}

A full statement of stellar {\it evolution} using entropy production and
variational principles is beyond the scope of this paper, and only a sketch is
presented in this section.   {\it Evolution} refers only to nuclear chemical 
evolution, not to faster hydrodynamic or thermal changes with fixed nuclear 
abundances (Kippenhahn and Weigert 1990; Arnett 1996).  The latter two 
processes are only time-dependent versions of the hydrostatic and thermal NESS.
Such dynamical but non-evolutionary behavior can be treated with time-dependent
extensions of the static variational principles stated in section~3.  The
hydrodynamic part is derived by minimizing the action (Lamb 1945; Sieniutycz
and Salamon 1990; Sieniutycz 1994):
\begin{equation}
\int dt \int dm\ [{\partial\Phi\over{\partial t}} - {1\over 2}
({\bf\nabla}\Phi )^2 +  e(\rho ,s)]\quad ,\eqnum{5.1}
\end{equation}\noindent
where $\Phi$ is the velocity potential, ${\bf v}$ = $-{\bf\nabla}\Phi$ =
the macroscopic velocity of fluid flow.  This action principle is adiabatic 
and holds only for irrotational fluid flow.  The isothermal analogue of~(5.1) 
can be obtained by substituting $f(\rho ,T)$ for $e(\rho ,s).$  
(A hydromechanical small-oscillation variational principle has been derived
for asteroseismology by Backus and Gilbert 1967).
The non-steady thermal behavior can be derived by minimizing the variational 
integral (Donnelly {\it et al.} 1996; Glansdorff and Prigogine 1971):
\begin{equation}
\int dt \int dV\ \Bigl[\sigma_{\rm cond}\ +\ \sigma_{\gamma\ {\rm diff}}\ +\ 
\sigma_{\rm conv} + {{\dot q}_*\over T} + [\rho c_VT^2]_*{\partial\over{
\partial t}}\Bigl({1\over T}\Bigr)\Bigr]\quad ,\eqnum{5.2}
\end{equation}\noindent
where $c_V$ is the specific heat capacity at constant volume, assuming the
mechanical part to be instantaneously in the NESS (Sieniutycz and Salamon 1990;
Sieniutycz 1994).  Equation~(5.2) is not a minimum entropy production 
principle, because it represents a NESS only if $\partial T/\partial t$ = 0.

The two variational principles~(5.1,2) can be be used to treat stellar
oscillations, with or without linearization.  In general, these principles
encompass only the propagation of seismic waves, not their driving forces,
unless the latter are explicitly included in the integrals.

\subsection{Gravitational Settling in Chemical Steady State}

Stellar evolution involves two fundamental changes: gravitational settling of 
heavier nuclei arising from spatially inhomogeneous nuclear fusion reactions 
and thermonuclear transmutation of elements.
In the Sun, for example, the associated global time scales
are $6\times10^{13}$ and $10^{10}$ yrs, respectively.  Both  evolution
processes can be described in terms of entropy production, but only
element diffusion is close to local chemical equilibrium (Kippenhahn and
Weigert 1990; Bahcall and Pinsonneault 1992).  As the mechanical
and thermal timescales are usually much shorter than the chemical timescale, 
the star is at each instant in evolution in both hydrostatic and thermal NESS.
But the boundary conditions for these NESS's change with chemical evolution, 
as the star's mass, radius, and luminosity change.

Chemical diffusion by gravitational settling can be dynamically formulated in 
terms of a minimum entropy production $\Sigma_{\rm nuc\ diff}$ (Donnelly {\it
et al.} 1966; Glansdorff and Prigogine 1971):
\begin{equation}
\begin{array}{ccc}
\Sigma_{\rm nuc\ diff} & = & \int dV\ \Biggl[\sum_{ij} {1\over 2}
[{\cal D}_{ij}(\rho ,T)]_*
{\bf\nabla}\Bigl({\mu_i-\mu_H\over T}\Bigr){\bf\nabla}\Bigl({\mu_j-\mu_H\over 
T}\Bigr) \nonumber \\
 &  & + \sum_i [{\cal D}_{iT}(\rho ,T)]_*{\bf\nabla}\Bigl({\mu_i-\mu_H\over T}
\Bigr){\bf\nabla}\Bigl({1\over T}\Bigr)\Biggr]\quad ,\end{array}
\eqnum{5.3}
\end{equation}\noindent
where the $\mu_i({\bf r}, t)$ are chemical potentials of the nuclear
species $i,$ and the ${\cal D}_{ij}(\rho ,T)$ are the chemical diffusion 
coefficients, including cross terms between different species.  The mixed
thermodiffusion effect is included with the ${\cal D}_{iT}$ terms.  By 
Onsager's
theorem, ${\cal D}_{ij} = {\cal D}_{ji}$ (Onsager 1931a, 1931b).  
Each chemical potential
is defined relative to some reference potential, here taken to be that
of hydrogen (H).
This form of chemical diffusion assumes a quasi-linear relation
between the element flux {\bf J}$_i$ and the chemical gradients ${\bf\nabla}
\mu_i:$
\begin{equation}
{\bf J}_i = - \sum_j {\cal D}_{ij}(\rho ,T) {\bf \nabla}\bigl({\mu_i-\mu_H
\over T}\bigr) - {\cal D}_{iT}(\rho ,T) {\bf \nabla}\bigl({1\over T}
\bigr)\quad ,\eqnum{5.4}
\end{equation}\noindent
in analogy with heat conduction.  The diffusion represented by~(5.3) is therefore in a NESS,
as equation~(5.4) is time-independent.

Chemical diffusion could also be formulated in terms of the extensive variables  
$N_i,$ the species densities $n_i,$ or the nuclear mass abundances $X_i,$
by use of the {\it grand potential:}
\begin{equation}
\begin{array}{ccc}
A(V,T,\mu_i) & = & F - \int dV\ \sum_i\mu_in_i\quad , \nonumber \\
n_i & = & -\rho (\partial a/\partial\mu_i)\quad ,\end{array}
\eqnum{5.5}
\end{equation}\noindent
where $a(\rho ,T,\mu_i)$ is the specific grand potential.

\subsection{Thermonuclear Burning as Nonlinear Reactions}

The entropy production of nuclear fusion is composed of a radiation part 
$\Sigma_\gamma ,$
already discussed in section~3, and a matter part $\Sigma_{\rm matter},$ 
hitherto ignored, because it is significant only in late stages of stellar 
evolution.
Starting with the Gibbs form of the entropy increment (De Groot and Mazur 1962;
Gyarmati 1970; Glansdorff and Prigogine 1971), the canonical 
expression for $\Sigma$ of matter undergoing chemical/nuclear reactions is:
\begin{equation}
\Sigma_{\rm matter} = \int dV\ \sum_i\ {{\dot n}_i\mu_i\over
T}\quad .\eqnum{5.6}
\end{equation}\noindent
where $n_i$ are the number densities of each species $i.$  
The photon chemical potential $\mu_\gamma$ is zero, because photon
number is not conserved.  The neutrino chemical potential $\mu_\nu$ is
zero as long as neutrinos have no significant density.

A more transparent form can be obtained by
defining common reaction rate densities ${\dot\xi}_\alpha$ = ${\dot 
n}_i/\nu_{\alpha , i},$ where the subscript $\alpha$ labels the reaction.
The $\nu_{\alpha , i}$ are the stoichiometric coefficients of reaction 
$\alpha :$
\begin{displaymath}
\nu_{\alpha ,1}N_1 + \nu_{\alpha ,2}N_2 + ...\rightarrow -\nu_{\alpha ,j+1}N_j
-\nu_{\alpha ,j+2}N_{j+2} + ...\quad ,
\end{displaymath}\noindent
for all participating reactants $i$ = 1...$j$ and products $i$ = $j+1$....
The {\it affinity} ${\cal A}_\alpha$ of reaction $\alpha :$
\begin{equation}
{\cal A}_\alpha = \sum_i \nu_{\alpha ,i}\mu_i\quad \eqnum{5.7}
\end{equation}\noindent
is a measure of how far a given reaction $\alpha$ is from chemical equilibrium.
For a dead star (one whose nuclear reactions have gone to completion), 
${\cal A}_\alpha$ = 0 for all $\alpha .$  The alternate form for $\Sigma_{\rm
matter}$ is then:
\begin{equation}
\Sigma_{\rm matter} = \int dV\ \sum_\alpha {{\cal A}_\alpha
{\dot\xi}_\alpha\over T}\quad ,\eqnum{5.8}
\end{equation}\noindent
a canonical bilinear form in the forces ${\cal A}_\alpha /T$ and fluxes 
${\dot\xi}_\alpha .$  

Because ${\cal A}_\alpha\sim$ MeV is higher than the ambient temperature $T,$
the thermodynamic forces driving nuclear burning are large.   This makes
stellar thermonuclear reactions highly nonlinear, even locally (Reichl 1980).
Because no general linear or quasi-linear relation holds between the fluxes
and the forces in the matter contribution to $\Sigma$,  without
additional knowledge or details, we can go no further in expressing
the entropy production of matter due to nuclear fusion than the kinematical
expressions~(5.6, 8).  Without a quadratic (or any other expression) for  
$\Sigma_{\rm matter}$  in terms of the forces ${\cal A}_\alpha /T$ there is no
automatic principle of minimum entropy production (Prigogine 1945; 
Lavenda 1978).

A variational principle of the quadratic type requires a quasi-linear relation between forces
and fluxes.  A non-quadratic minimum entropy production principle
for matter fusion may still be possible, if the chemical evolution is in a NESS, with the fluxes
${\dot\xi}_\alpha$ constant in time.  This NESS approximation holds as long as the star
remains on one branch of hydrostatic thermonuclear burning (H,
He, C, Ne burning), or on one of the late, hydrodynamic stages of chemical evolution
(O, Si burning) (Arnett 1996).  Furthermore, each burning stage constitutes a 
different
NESS.  It is possible that minimum $\Sigma_{\rm matter}$ holds on
each burning stage, but that the bifurcation from one burning stage to the
next is a discontinuous phase transition similar to the switch from 
non-convective to convective heat transport (Appendix~B).  So long as some 
fuel remains locally from one stage, that burning continues; the minimum 
entropy production
conjecture would imply that only when its fuel runs out is the next burning
stage preferred on entropy production grounds.
Similar considerations may also hold for the evolution of protostars.

If the matter fusion is not representable as a NESS, another, time-dependent 
variational
principle, analogous to~(5.2), may still be possible, lacking a simple 
interpretation as minimum entropy production.

\section{Summary and Outlook}

Variational principles are most useful in expressing general properties of
static or dynamical systems, such as exact or approximate symmetries, and in 
suggesting generalized coordinates.  Integral principles also open alternative 
paths to solution via an iteration of approximate solutions.  Under some 
restrictions, such methods converge and provide a useful replacement for 
numerical differential methods (Donnelly {\it et al.} 1966).

We have presented two different pairs of variational principles to replace the
standard four equations of a hydrostatic NESS and steady luminosity flow.
Two of these, the thermal variational principles~(3.6-8) and~(4.4) for
NESS luminosity transport, are the principal original results of this paper. 
These represent a global alternative to numerical integration of
the four differential equations of stellar structure.  Practically, these 
thermal variational principles suggest calculational procedures analogous to
either the relaxation technique or the Rayleigh-Ritz method (Donnelly {\it et 
al.} 1966; Glansdorff and Prigogine 1971).  Examples of global analytic
approximations will be published elsewhere.

As seen in section~5, the static variational integrals of sections~3 
and~4 can be extended to the time-dependent hydrodynamic and thermal 
non-steady states.

We have been able to recast stellar evolution only partially into integral 
form, and a dynamical principle is lacking.  A complete integral reformulation
of stellar theory would encompass these outstanding issues: the transmutation 
of elements, the emission of neutrinos, and bifurcations and instabilities
during the multiple burning stages.

\acknowledgments

We are indebted to Christopher Essex (Univ. W. Ontario)
for many helpful discussions and suggestions.  
This research was supported at the Institute for Theoretical
Physics, University of California, Santa Barbara, by the National Science 
Foundation under Grant No. PHY89-04035; by the University of Florida, Institute
for Fundamental Theory of the Department of Physics; by the Aspen Center
for Physics; by the Telluride Summer Research Center and Telluride Academy; 
and by the Department of Energy under Grant Nos. DE-FG05-86-ER40272 (Florida) 
and DE-AC02-76-ERO-3071 (Penn).
We thank the ITP, the Aspen Center, and the TSRC for their hospitality.

\appendix\section*{Appendix A: Conductive and Radiative Entropy Production}

Let us define notation and simple concepts first.
LTE is assumed to be valid in the strictly local limit,
with small deviations over small distances.  Thus matter and radiation are
in the maximal entropy state locally, with non-zero second-order
thermodynamic fluctuations about the LTE.  Purely local state variables
and functions carry an asterisk (*) subscript.  These are not subject to
the non-equilibrium functional variations (Donnelly {\it et al.} 1966;
Glansdorff and Prigogine 1971; Chandrasekhar 1939 and 1950; Essex 1984a,
1984b; Holden and Essex 1996).

\subsection*{1. Heat Conduction}

The familiar case of heat conduction best introduces
the entropy production $\Sigma .$  This function's spatial density
$\sigma$ takes on, for matter alone, a simple bilinear form derived from
the Gibbs equilibrium formula for the entropy increment $dS$ = $dQ/T$ +
{\it mechanical, chemical,... etc. terms} (Callen 1985; Balian 1992).
The entropy produced
by a heat current {\bf J}$_Q$ flowing through a temperature field $T$ in a
volume $V$ with a surface area $A$ is:
\begin{equation}
\begin{array}{ccc}
\Sigma_{\rm cond} & = & \int d{\bf A}\cdot ({\bf J}_Q/T)\quad \nonumber \\
       & = & \int dV\ {\bf\nabla}\cdot ({\bf J}_Q/T)\quad .\end{array}
\eqnum{A.1}
\end{equation}\noindent
In a steady state, $\nabla\cdot{\bf J}_Q$ = $\dot q,$ where $\dot q$
is the heat source density within the volume.  On the other hand, {\bf J}$_Q$
in the quasi-linear case is linear in an externally given inverse-temperature
gradient, with:
\begin{equation}
{\bf J}_Q = (kT^2){\bf\nabla}(1/T)\quad ,\eqnum{A.2}
\end{equation}\noindent
where $k(\rho ,T)$ is the local thermal conductivity, an integral over the
microscopic momentum space:
\begin{equation}
kT^2 = (4\pi /3)\int^\infty_0 dp\ p^2\ f_0({\bf r}, {\bf p})
[v^2(\rho e)^2/w(p)]\quad ,\eqnum{A.3}
\end{equation}\noindent
where $v$ and $p$ are the atomic speed and momentum, $\rho e$ the gas internal 
energy density,
$f_0$ the reduced 1-particle phase space distribution at LTE, and
$w(p)$ the collision time (Balian 1992).  Thus:
\begin{equation}
\sigma_{\rm cond} = (kT^2)_*[{\bf\nabla}(1/T)]^2 + {\dot q}_*/T
\eqnum{A.4}
\end{equation}\noindent
is the entropy density production rate by the transport and 
production  of heat.

In~(A.4) the current {\bf J}$_Q$ must be evaluated
assuming a {\it given} external $T$ gradient.  In reality, this gradient
is produced by the current itself.  Thermodynamically, the form~(A.4) assumes
an externally-imposed {\it first-order} deviation in entropy from LTE, while
in fact, the deviation for the transport term is a {\it second-order} 
fluctuation (Donnelly {\it et al.} 1966; Glansdorff and Prigogine 1971). 
(LTE makes the first-order fluctuations
of entropy zero.)  Correction of this problem leads to
an extra factor of 1/2 in the first (transport) term.  Note that this
term is quadratic in the gradient.  The second (production) term receives
no extra factor of (1/2), as it really is a first-order deviation in entropy:
the heat production ${\dot q}$ is an externally given function.  Fourier's 
equation
for heat conduction is obtained by varying $T$ in
\begin{equation}
\Sigma_{\rm cond} = \int dV\ (1/2)(kT^2)_*[{\bf\nabla}(1/T)] + 
{\dot q}_*/T]\quad ,\eqnum{A.5}
\end{equation}\noindent
holding $(kT^2)_*$ and ${\dot q}_*$ fixed.  Afterwards, $T_*$ is 
self-consistently 
set equal to $T.$   The heat flux is
\begin{displaymath}
{\bf J}_Q = (l/4\pi r^2){\bf\widehat r} = -k{\bf\nabla}T\quad .
\end{displaymath}\noindent

\subsection*{2. Radiative Diffusion}

Radiative transport by photon diffusion is formally similar to heat conduction 
by matter-matter collisions (Chandrasekhar 1950; Essex 1984a, 1984b;
see also: Planck 1913; Rosen 1954; Kr\" oll 1967; Lallemand and Martinet 1979).
The role of $k$ is taken by an expression
involving the opacity $\kappa$ of the matter to photon travel.   We
should then expect $\kappa$ to involve an integral over the {\it photon}
phase space.  The evaluation of $\Sigma_\gamma$ for LTE with a small
gradient begins with the {\it generalized} bilinear form for photons at
angular frequency $\omega$ passing through and interacting with matter at 
temperature $T:$
\begin{equation}
\sigma_{\gamma\ {\rm diff}} = 2\pi\int^\infty_0 d\omega\ \int^{+1}_{-1} 
d\xi\ J_\omega[1/T_\omega - 1/T]\quad ,\eqnum{A.6}
\end{equation}\noindent
where $\xi$ = photon local direction cosine ( not to be confused with the 
reaction rate densities 
$\dot\xi_\alpha$ of section~5), 
 $J_\omega$ is the
differential radiation luminosity density out of equilibrium:
\begin{equation}
J_\omega = \kappa_\omega\rho [B_\omega - I_\omega ]\quad ,\eqnum{A.7}
\end{equation}\noindent
with $\kappa_\omega$ the frequency-specific opacity of matter, $B_\omega$ the
Planck function (blackbody differential radiation energy flux), and
$I_\omega$ the true energy
flux of photons.  In the spherical diffusion approximation,
$I_\omega$ = $B_\omega$ -- $(\xi /\kappa_\omega\rho )
(\partial B_\omega /\partial r)$ 
with the gradient term small  except very near the stellar surface.  
$T_\omega$ is the
effective {\it brightness temperature} for any $I_\omega$ and varies with
$\omega :$
\begin{equation}
1/T_\omega = (1/\hbar\omega )\ln\Bigl[{2\hbar\omega^3\over{8\pi^3c^2I_\omega}}
 + 1\Bigr]
\quad .\eqnum{A.8}
\end{equation}\noindent
In the {\it Rosseland mean opacity:}
\begin{equation}
{1/\kappa }\equiv \Bigl(\int^\infty_0 d\omega\ [1/\kappa_\omega]\ \partial 
B_\omega /
\partial T\Bigr) /\Bigl(\int^\infty_0 d\omega\ \partial B_\omega /\partial T
\Bigr)
\quad ,\eqnum{A.9}
\end{equation}\noindent
the denominator has the value $acT^3.$

The entropy production is a generalization of the bilinear Gibbs formula, with
changes in $1/T$ playing the role of the thermodynamic force (which is
functionally varied) multiplying some flux, as outlined in Section~3.

\subsection*{3. Source and Boundary Terms}

Nuclear burning produces entropy by the 
production of high-energy photons and the thermalization  of fusion products.
The radiation/kinetic energy is produced in a fusion reaction by thermalized
matter, a tiny, positive contribution
to radiation entropy, as the original matter reactants are thermalized.
This original photon/kinetic energy
is absorbed upon thermalization, a negative contribution to
entropy.  Both the matter kinetic energy and radiation are then
thermalized to the
ambient temperature of the core, a large and positive contribution.
The first two contributions are negligible compared to the third, being
suppressed by the ratio $T/T_0,$ where $T_0$ is the brightness temperature
of the original photons, in the range 0.1-5 MeV, well above the typical
stellar core temperature.  These contribution are small but non-negligible
for older stars with higher core temperatures.

For any kind of radiation transport, there is a constant
entropy production  from the release of radiation into empty space.
The local radiation entropy flux {\bf H} has magnitude $(4/3)(ac/4)T^3_{\rm
eff}$ on a stellar surface with temperature $T_{\rm eff}.$  Multiplying
this expression by the surface area $4\pi R^2$ gives the total boundary
entropy production, $\Sigma_{\rm bound}.$
As in conduction, the transport term arises from second-order
fluctuations in the LTE entropy; thus the factor of 1/2 again.
The luminosity source term $\varepsilon\rho$ is again a true external
first-order deviation from LTE, so the complete entropy production is:
\begin{equation}
\begin{array}{ccl}
\Sigma_\gamma & = & \Sigma_{\gamma\ {\rm diff}} + \Sigma_{\rm fusion} \nonumber \\
              & = & \int dV\ \Bigl[ (1/2)[4acT^5/(3\kappa\rho )]_*
  [{\bf\nabla}(1/T)]^2 + (\varepsilon\rho)_*/T\Bigr]\quad ,\end{array}
\eqnum{A.10}
\end{equation}\noindent
in the radiative diffusion case, where $T_*$ and $M$ are not varied.

A more realistic procedure would be to construct the $\Sigma_\gamma$ for
the stellar atmosphere.  Multiple matter temperatures and a variable
radiation brightness temperature must then be introduced.  We do not
discuss stellar atmospheres in further detail.

A sense of the relative sizes of these entropy production rates may be
gotten from the example of the Sun:
Conductive heat transport is small; the core and
radiative zone transport luminosity by radiative diffusion.
We estimate the various entropy production rates, using
the boundary contribution $\Sigma_{\rm bound}\sim$ 
8$\times$10$^{29}$ erg/$\K$ sec as a benchmark.
The radiation entropy production in the core is much smaller, because of
the much higher temperature $T_{\rm eff}\simeq$ 5500$\K$ versus $T_c\simeq
15\times 10^6 \K :$ $\Sigma_{\rm fusion}\simeq$
(0.0007)$\Sigma_{\rm bound}.$  The entropy production rate due to 
transport of the luminosity from the core to the surface is $\Sigma_{\rm diff}
\simeq$ (0.0002)$\Sigma_{\rm bound}.$  This
contribution is also much smaller than the boundary term, but {\it not}
much smaller than the core term.  Essentially all of the
radiation entropy production in the Sun arises from the release of
radiation into empty space at the surface.  But, since this term  arises from 
boundary conditions, it is not varied in deriving the transport equation.

\appendix\section*{Appendix B: Convective Entropy Production}

Unlike conductive and radiative transport, convection is a macroscopic process, 
involving bulk motion of matter.  Furthermore, real convection is complicated,
with many length scales and the possibility of significant
turbulence effects.  No satisfactory theory of convection exists, but
there are useful models that capture the essentials of the heat transport,
enough for consideration in the NESS structure of MS stars (Chandrasekhar 
1939; Lamb 1945; Donnelly {\it et al.} 1966; Glansdorff and Prigogine 1971; 
Stix 1989; Sieniutycz and Berry 1989, 1991, 1992, 1993; Hansen and Kawaler 
1994).

\subsection*{1. Schwarzschild Criterion and Mixing Length Theory}

Schwarzschild's well-known picture
begins by idealizing the convective bulk motion of matter cells slightly
hotter than their surroundings and heated from below in a gravitational
field by a temperature gradient.  In the basic {\it mixing length theory}
(MLT), such a cell, with linear size $\mixl ,$
moves upwards a distance $\mixl$ before breaking
up and merging with the surrounding gas or fluid, releasing its heat in
the process.  (In real convection, there are many mixing scales, not one.)
The cell floats upwards by bouyancy, its internal temperature $T^\prime$
being higher and density $\rho^\prime$ lower than its surroundings.  In MLT
(Boussinesq approximation), the effects of sound
and shock waves are ignored; the pressures internal and external to the
cell are assumed equal; and the $T$ and $\rho$ variations
between the inside and outside of the cell are small: $(\rho - \rho^\prime )/
\rho$ and $(T^\prime - T)/T\ll$ 1.  In stellar
applications, in addition, $\mixl$ is assumed much smaller
than the stellar radius $R$ (Hansen and Kawaler 1994).

In order for convection to occur at all, the temperature contrast between
the bottom of a mixing length, where the cell is heated, and the top must
be large
enough to overcome gravity by creating a bouyancy force. In the continuum, 
this leads to Schwarzschild's criterion for convection to occur:
\begin{equation}
\nabla > \nabla_{\rm ad}\quad ,\eqnum{B.1}
\end{equation}\noindent
where $\nabla\equiv -d\ln T/d\ln P,$ 
the temperature $T(r)$ and pressure $P(r)$ are the actual profiles of
a given star.  $\nabla_{\rm ad}$ is the same as $\nabla ,$ evaluated
for the same star, but as if the equation of state were adiabatic,
which assumes that the cell exchanges heat at most very slowly with its
surroundings before it breaks up. 
In the convective regime, the temperature gradient in the star outwards is 
steep
enough that relatively hotter cells are bouyant.  The simplicity of this
condition hides an important complication: the presence or absence of
convection itself changes the $T(r)$ and $P(r)$ profiles.  In practice, the
problem is solved iteratively and self-consistently.  For {\it
quasi-adiabatic} convection $\nabla$ is only slightly larger than $\nabla_{\rm
ad},$ and the details of this ``slow but hot'' convection are unimportant.  
For $\nabla$ well above $\nabla_{\rm ad},$ the specific
assumptions of the convection model become important, including 
turbulence in stellar conditions with very low viscosity (very high Reynolds 
number) and smallness of $\mixl$
compared to the pressure scale height $\lambda_P.$  Such convective zones are 
then ``warm but fast''; but, typically being in the outer parts of stars, they
are almost irrelevant to the stellar cores.

\subsection*{2. Entropy Production: Heat Loss and Bouyancy}

As a convective cell rises, it can lose part of its excess heat by
thermal conduction/radiative diffusion (the two can be treated on an
equal footing by using the equivalence of Appendix~A) and by viscosity.
The total entropy production of convection $\Sigma_{\rm conv}$ is a sum
of three terms: the first due to heat loss (positive), the second to
viscosity (positive), and the third to bouyancy (negative).  The last is
negative because bouyancy converts heat to the work of raising the
cell in the local gravitational field ``g''.
\begin{equation}
\begin{array}{ccl}
\sigma_{\rm conv} & = & \sigma_{\rm heat\ loss} + \sigma_{\rm visc} + \sigma_{\rm
bouy}\quad , \nonumber \\
\sigma_{\rm heat\ loss} & = & [(\rho c_P)^2/ (2k)]
  \Bigl[w\mixl /\lambda_P\Bigr]^2~(\nabla - \nabla_{\rm ad})^2\quad ,\\
\nonumber
\sigma_{\rm visc} & = & \eta w^2/(T\mixl^2)\quad , \nonumber \\
\sigma_{\rm bouy} & = & -[2(\rho {\rm g}\gamma_P)(w\mixl )^2/(\eta_T
  T\lambda_P)]~(\nabla - \nabla_{\rm ad})\quad ,\end{array}
\eqnum{B.2}
\end{equation}\noindent
where $w$ is the upward velocity of the cell.  In MLT, $w$ is solved for
self-consistently: 
\begin{displaymath}
w = \mixl\sqrt{(\gamma_P{\rm g}/\lambda_P)(\nabla - \nabla_{\rm ad})}\quad .
\end{displaymath}\noindent  
The kinematic viscosity is $\eta$, 
the  thermal diffusivity $\eta_T\equiv k/\rho c_P,$ and $c_P$ is the 
specific heat capacity of matter at constant pressure.  The isobaric exponent
$\gamma_P\equiv$ $-(d\ln\rho /d\ln T)_P.$
The temperature contrast of the cell $(T^\prime)$ with its
surroundings $(T)$ is 
\begin{equation}
{1\over T} - {1\over{T^\prime}} = {1\over T}\Bigl({w\over\lambda_P}\Bigr)
\Bigl({\mixl^2
\over \eta_T}\Bigr) [\nabla - \nabla_{\rm ad}]\quad .\eqnum{B.3}
\end{equation}\noindent 
The convection is {\it efficient} if the cell's cooling time $\mixl^2/\eta_T$ 
is long compared to its travel time $\lambda_P/w.$  In stars, the
viscosity is still smaller and negligible in MLT, i.e., $\eta\ll\eta_T.$

The inverse temperature difference that acts as the thermodynamic force for
heat transfer is just the expression~(B.3).  The relative temperature contrast,
$T$ times equation~(B.3), must be small for quasi-linear thermodynamics to
be valid.  The range of its validity is identical to that of MLT.  In the
quasi-adiabatic or ``slow but hot'' regime, $\nabla - \nabla_{\rm ad}$ is 
positive
but so tiny that the relative temperature contrast is small.  In the ``warm
but fast'' regime, $\nabla - \nabla_{\rm ad}$ is positive and not tiny, and the
relative temperature contrast is not small.  Quasi-linearity and MLT are then
not valid, although MLT is commonly used anyway, for the lack of a better but
still simple model.  The specific details of convection also matter in this
regime, and the simplicity of the quasi-linear situation vanishes.

Convection and convective instability are illuminated by comparing the 
competing entropy production rates for 
the {\it same cell} in two regimes, convective ({\it heat loss} + {\it 
bouyancy}) and
non-convective ({\it heat loss} alone) (Figure~1) (Donnelly {\it et al.} 1966;
Glansdorff and Prigogine 1971).  Let us
start by taking a convective motion with an arbitrary cell velocity $w.$
Without the bouyancy part, the entropy production due to heat conducted
out of the cell falls and then rises quadratically, reaching zero at $\nabla$ 
= $\nabla_{\rm ad}$ (Figure~1: solid line).  (The cell velocity is actually
zero in this case and must be replaced by an equivalent expression for pure
conduction.)  If we put the bouyancy part
back in, the entropy production curve is modified by a negative term linear
in $\nabla - \nabla_{\rm ad}$ and becomes distorted (Figure~1: dashed line).
The two curves cross at zero, for $\nabla$ = $\nabla_{\rm ad}.$  The branch 
with
lower entropy production for $\nabla < \nabla_{\rm ad}$ is the {\it heat loss}
alone.  But for $\nabla > \nabla_{\rm ad},$ the {\it heat loss + bouyancy}
branch has the lower entropy production rate and is favored.  In this regime,
the cell becomes bouyant, and convection begins.  The analogy with first-order
equilibrium phase transitions is evident, with the entropy production rate
taking the place of the free energy.

In the case of convection, there is no variational calculation for the
thermal structure, only a
comparison of two discrete branches of $\Sigma .$  The cell velocity $w$
must be computed from hydrodynamic considerations, a topic beyond the
scope of this paper.
\placefigure{FIG1}

\appendix\section*{Appendix C: Neutrino Entropy Production}

Because untrapped neutrinos are not in LTE, they do not contribute directly to stellar
structure. While the total lepton number (charged leptons plus neutrinos) is
conserved, neutrino number is generally not conserved.  After their production
in the hot stellar core, in quasi-stellar stars, neutrinos suffer negligible weak
interactions and change in neutrino number.  They therefore transport both
neutrino luminosity and number, while electromagnetic radiation transports only
photon luminosity.
Neutrinos contribute to quasi-static stellar structure only indirectly by
cooling of matter.

\subsection*{1. Neutrino Number, Energy, and Entropy Production}

In analogy with the specific photon/heat luminosity production rate,
a specific neutrino rate, $\varepsilon_\nu ,$ is defined.  As 
for photons, 
this quantity is a sum over all neutrino-producing reactions, 
\begin{equation}
\rho\varepsilon_\nu = \int dV \int dE\ E\ {\dot n}_E\quad ,\eqnum{C.1}
\end{equation}\noindent
where ${\dot n}_E$ is the neutrino differential production rate density and
we label the neutrino phase space by neutrino energy $E,$ not 
frequency.
The {\it total} specific luminosity is the function
$\varepsilon .$
But unlike photons, neutrinos are not thermalized after they are created,
leaving the star unimpeded.  Their only memory of thermodynamics is
the thermalized matter that produced them.  In analogy with the photon entropy
production (see Appendix A), the neutrino entropy production may be
written:
\begin{equation}
\Sigma_\nu = \int dV \int dE\ {E\ {\dot n}_E\over T_E}\quad ,\eqnum{C.2}
\end{equation}\noindent
where ${\dot n}_E$ is the neutrino differential production rate density and
we integrate over the single neutrino energy $E,$ instead of frequency.
Unlike photons, there is no matter heat bath term for free-streaming 
neutrinos.
Technically, there is a second term in~(C.2) giving the tiny contribution to
$\Sigma_\nu$ from the neutrino absorption (as in terrestrial detectors).
The overall structure of 
$\Sigma_\nu$ is non-local because neutrinos are generally not in LTE.

Neutrinos have an energy-dependent brightness temperature $T_E,$ but, if
unconfined, no chemical potential.
This temperature is defined by the Fermi-Dirac analogue of (A.8):
\begin{equation}
{1\over T_E} = {1\over E}\ln\Bigl[{2(E/\hbar c)^3c\over{8\pi^3 I_E}} - 1
\Bigr]\quad ,\eqnum{C.3}
\end{equation}\noindent
where $I_E$ = $E\ N_E$ is the energy-specific neutrino differential energy 
flux; $N_E$ is the same for neutrino number.  Define additional functions:
\begin{equation}
\begin{array}{c}
{\dot n}_\nu = \int dE\ {\dot n}_E\quad ,\quad {\dot N}_\nu = 
\int dV\ {\dot n}_\nu\quad , \nonumber \\
{\dot N}_E = \int dV\ {\dot n}_E\quad ,\quad N_E = {\dot N}_E/4\pi r^2\quad ,
\end{array}\eqnum{C.4}
\end{equation}\noindent
where the last definition holds only in the case of spherical symmetry.
The exact result of~(C.4) depends on the specific reaction.  (Neutrino entropy
production will be further discussed in future publications; see Essex and
Kennedy 1996.)

\subsection*{2. Examples}

For an infinitely sharp line, unbroadened by thermal fluctuations, the
entropy production rate is zero: the neutrinos are produced at exactly one
energy.  The effect of thermal broadening for neutrino line emission can be 
treated in the same way as it is for photon emission lines (Stix 1989).  Define
the thermally broadened shape function $\phi_E(E):$
\begin{equation}
\phi_E(E) \equiv {H(b,z)\over{\sqrt{\pi}{\Delta E}_D}}\quad ,\eqnum{C.5}
\end{equation}\noindent
where $H(b,z)$ is the Voigt function:
\begin{equation}
\begin{array}{c}
H(b,z) \equiv (b/\pi )\int^\infty_{-\infty} 
dy\ \exp(-y^2)/[(z-y)^2+b^2]\quad , \nonumber \\
\int^\infty_{-\infty} dz\ H(b,z) = \sqrt{\pi}\quad .\end{array}
\eqnum{C.6}
\end{equation}\noindent
$I_E$ and $N_E$ are proportional to $\phi_E.$
The reduced variable $z\equiv (E - E_0)/{\Delta E}_D,$ where $E_0$ is the 
center of the neutrino line and ${\Delta E}_D$ is the Doppler width:
\begin{equation}
{\Delta E}_D \equiv E_0\sqrt{{2T\over{m_Ac^2}} + {\xi^2_t\over c^2}}\quad .
\eqnum{C.7}
\end{equation}\noindent
$A$ is the nucleus emitting the neutrino, and $\xi_t$ is the root-mean-square 
{\it microturbulence} speed, set to zero here for simplicity.
Define $\gamma$ as twice the effective collision rate for the nucleus $A$ in
the plasma; then:
\begin{equation}
b \equiv {\hbar\gamma\over{4\pi{\Delta E}_D}}\quad .\eqnum{C.8}
\end{equation}\noindent
The collision rate is dominated by collisions of $A$ with the plasma electrons,
and, usually, $b\ll$ 1.  If $b\ll$ 1, $H(b,0)$ = 1 + ${\cal O}(b).$  

In the spherical case,
\begin{equation}
\Sigma_\nu = \int dV \int dE\ [{\dot n}_E]_*\ \ln\Bigl[{(E/\hbar c)^2\ r^2\over
{\pi^2\hbar\ {\dot N}_E}} - 1\Bigr]\quad ,\eqnum{C.9}
\end{equation}\noindent
where the background functions are distinguished again by *.  This expression
is non-local, since we lack neutrino LTE: 
\begin{displaymath}
{\dot N}_E = \int dV^\prime\ {\dot n}_E(r^\prime )\quad .
\end{displaymath}\noindent
A crude estimate of $\Sigma_\nu$ for a broadened line is:
\begin{equation}
\Sigma_\nu \sim {\dot N}_\nu\ \ln\Bigl[{(E_0/\hbar c)^3R^2_c v_D\over{\pi^{3/2}
{\dot N}_\nu}}\Bigr]\quad ,\eqnum{C.10}
\end{equation}\noindent
where $v_D$ is the Doppler root-mean-square speed: $v_D$ = $c{\Delta E}_D/E_0,$
and $R_c$ is the radius of the neutrino-producing stellar core.  In the
Sun, for example, the neutrino lines produced by ${}^7Be$ contribute 
approximately 5$\times$10$^{18}$ erg/$\K$ sec to $\Sigma_\nu .$

If we let $T\rightarrow 0$ holding all other state variables fixed, the
entropy production $\Sigma_\nu$ of a line vanishes as $T^{1/2}\ln T^{1/2},$
as it should: at zero temperature, the reactant matter is not thermalized.

In the continuum case, we distinguish the actual neutrino energy $E$ from
the zero-temperature neutrino energy $E_0,$ which is now continuous.
We then define a double differential neutrino production rate density
${\dot n}_{EE_0},$ per unit $dE$ and per unit $dE_0.$  Then:
\begin{equation}
\Sigma_\nu = \int dV \int dE_0 \int dE\ [{\dot n}_{EE_0}]_*\ \ln\Bigl[
{(E/\hbar c)^2r^2\over{\pi^2\hbar{\dot N}_{EE_0}}} - 1\Bigr]\quad .
\eqnum{C.11}
\end{equation}\noindent
In the case of the Sun's dominant $pp$-produced neutrinos, we estimate a
contribution to $\Sigma_\nu$ of 10$^{23}$ erg/$\K$ sec.  Because the Sun's
photon luminosity so much dominates its neutrino luminosity, the solar 
neutrino entropy production rate is far smaller than that of solar photons.

\newpage
\figcaption{
Entropy production rate $\Sigma$ of a potentially convective cell conducting 
heat only (solid line) versus conducting heat and rising by convection 
together (dashed line).  Convective solution is preferred (lower $\Sigma$) if 
Schwarzschild criterion is satisfied: $\nabla > \nabla_{\rm ad}.$\label{FIG1}}

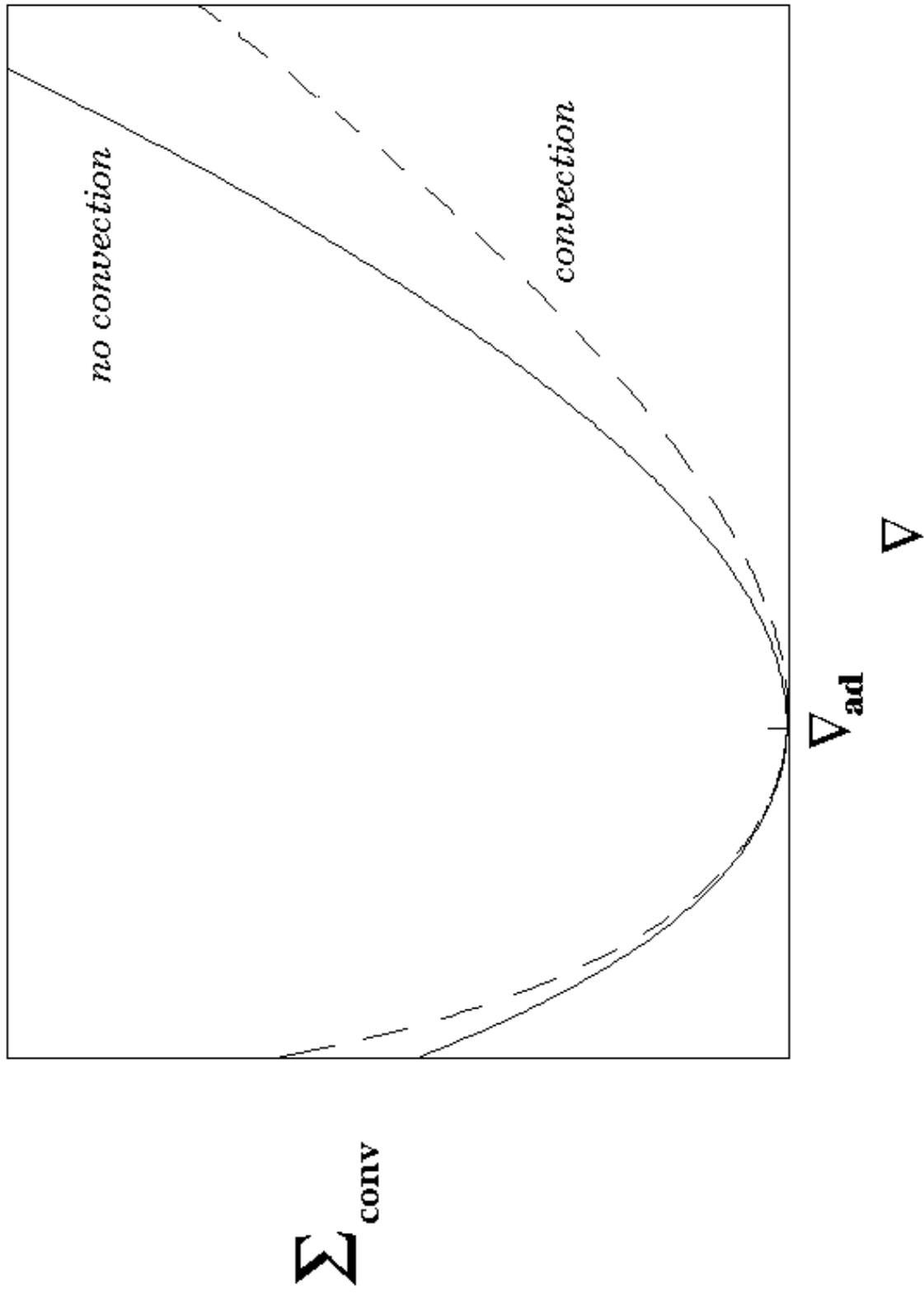
\end{document}

\begin{flushright}
UF-IFT-HEP-95-4\\ UPR-718-T\\ NSF-ITP-95-71\\ 
August 1997
\end{flushright}
\title{ERRATUM:\\ Variational Principles\\ for Stellar Structure}
\begin{center}
{\it Astrophys. J.} {\bf 484}, 329 (1997)
\end{center}
\author{Dallas C. Kennedy}
\affil{University of Florida}
\authoraddr{Gainesville FL 32611 USA;\\ kennedy@phys.ufl.edu}
\and
\author{Sidney A. Bludman}
\affil{University of Pennsylvania}
\authoraddr{Philadelphia PA 19104 USA;\\ bludman@bludman.hep.upenn.edu}


The photon entropy production $\Sigma_\gamma$ is an extensive quantity and thus
is fundamentally a volume integral of its local density $\sigma_\gamma .$
Re-expression of $\Sigma_\gamma$ as a surface integral is effected through
the divergence theorem and the surface photon entropy flux {\bf H}.  This step
requires the photons to propagate freely from a well-defined surface.  Below
this surface, the photons diffuse through matter instead.

In this case, of a star with a sharp surface of temperature $T_{\rm eff}$ and
no atmosphere, the paragraphs surrounding
equations~(3.10) and~(A.10) incorrectly
present the surface integral form of $\Sigma_\gamma$ (``the total boundary
entropy production, $\Sigma_{\rm bound}$'') as an {\it additional} contribution
to $\Sigma_\gamma$.  In fact the boundary expression is simply the {\it
total} $\Sigma_\gamma$ within the enclosed volume, {\it including} all
entropy production due to luminosity production and transport.  The volume
integrals~(3.9) and~(A.10) are the photon entropy production {\it up to, but
not including,} the surface; the surface integral or boundary term~(3.10) is
the photon entropy production including the surface, as well as the interior.

Thus, the contribution of the surface {\it alone} to $\Sigma_\gamma$ is the
{\it difference} of these two expressions: $\Sigma_{\rm surface}$ = 
$\Sigma_{\rm bound} -$ $\Sigma_{\rm volume}.$  The use of the volume 
integral~(3.9) alone in the variational principle~(3.11) for thermal structure
continues to be justified by the boundary conditions that hold the radius $R$ 
and $T_{\rm eff}$ fixed.